\documentclass{ws-procs9x6}

\usepackage{graphicx}
\newcommand{\pt}{p_\mathrm{T}}

\begin{document}

\title{Performance Study for a Muon Forward Tracker\\in the ALICE Experiment}

\author{A.~Uras$^*$ on behalf of the ALICE MFT Working Group}

\address{IPNL, Universit\'e Claude Bernard Lyon-I and CNRS-IN2P3,\\
Villeurbanne, France\\
$^*$E-mail: antonio.uras@cern.ch}


\begin{abstract}
ALICE is the experiment dedicated to the study of the quark gluon plasma in heavy-ion collisions at the CERN LHC. Improvements of ALICE sub‐detectors are envisaged for the upgrade plans of year 2017. The Muon Forward Tracker (MFT) is a proposal in view of this upgrade, motivated both by the possibility to increase the physics potential of the muon spectrometer and to allow new measurements of general interest for the whole ALICE physics. In order to evaluate the feasibility of this upgrade, a detailed simulation of the MFT setup is being performed within the AliRoot framework, with emphasis on the tracking capabilities as a function of the number, position and size of the pixel planes, and the corresponding physics performances. In this report, we present preliminary results on the MFT performances in a low-multiplicity environment. 
\end{abstract}


\bodymatter

\section{Introduction: Current Muon Arm Setup and Physics}

Lepton-pair measurements always played a key-role in high energy nuclear physics: leptons arrive to the detectors almost unaffected, being not sensible to the strong color field dominating inside the QCD matter, thus providing an ideal tool to probe the whole evolution of nuclear collisions. The \mbox{ALICE} experiment at the CERN LHC, representing the most recent effort in this field, measures single lepton and lepton pair production both at central rapidity, in the electron channel, and at forward rapidity, in the muon channel. 

~\\
Identification and measurement of muons in ALICE are performed in the Muon Arm\cite{Aamodt:2008zz}, covering the pseudo-rapidity region $-4 < \eta < -2.5$\footnote{~Being the collision symmetrical, we indicate the muon spectrometer acceptance using positive values in the following, both for the (pseudo-)rapidity and the $z$ coordinate.}. Starting from the nominal interaction point (IP) the Muon Arm is composed of the following elements. (i) A hadron absorber made of carbon, concrete and steel, between $z=0.9$ and $z=5.03$~m; its material budget corresponding to ten hadronic interaction lengths, it provides a reliable muon identification. (ii) A dipole magnet 5~m long providing a magnetic field of up to 0.7~T in the horizontal direction, corresponding to a field integral of 3~Tm. (iii) A set of five tracking stations, each one composed of two cathode pad chambers with a space resolution of about $\sim 100~\mu$m in the bending direction: the stations are located between $z=5.2$ and $z=14.4$~m, the first two ones upstream of the dipole magnet, the third one in its gap and the last two ones downstream. (iv) A 1.2 m thick iron wall, corresponding to 7.2 hadronic interaction lengths, placed between the tracking and trigger systems, which absorbs the residual secondary hadrons emerging from the front absorber. (v) The muon trigger system, consisting of two detector stations, placed at $z=16.1$ and $=17.1$~m, respectively, each one composed of two planes of resistive plate chambers, with a time resolution of about 2~ns. 

~\\
The ALICE experiment has already an intense physics program based on muon measurements, since the very start of its data taking. Within this program, currently active both in p--p and Pb--Pb collisions, three main directions can be identified: study of quarkonia production\cite{MartinezGarcia:2011nf}, of open Heavy Flavors (HF) production\cite{Dainese:2011vb}, of low mass dimuons\cite{DeFalco:2011wj}. The study of quarkonia production in p--p collisions allows one to investigate perturbative and non-perturbative aspects of QCD, by means of the analysis of the kinematic distributions of the resonances\cite{Aamodt:2011gj}; quarkonia production in nuclear collisions, on the other hand, is of primary importance to test quarkonia suppression/recombination mechanisms possibly being effective in deconfined QCD matter and already intensively investigated at the SPS and RHIC heavy-ion facilities. Open charm and beauty production, besides giving information on the initial stage of the nuclear collisions thanks to the short formation time of the heavy quarks, also provides sensitivity to the energy density of the deconfined matter through the mechanism of in-medium energy loss of heavy quarks. Low mass dimuons, finally, provide insight to soft QCD processes in the LHC energy regime, and allow one to characterize the properties of the deconfined medium through the analysis of the properties and the production cross sections of the light vector mesons.
%

\section{The Muon Forward Tracker Proposal}

The current ALICE muon physics program suffers from several limitations, basically because of the multiple scattering induced on the muon tracks by the hadron absorber. The details of the vertex region are then completely smeared out: in particular, this prevents us to disentangle prompt and displaced $J/\psi$ production (the production of $J/\psi$ from $b$ accounting for $\sim 20\,\%$ of the prompt cross section) as well as to disentangle open charm and open beauty without making assumptions relying on physics models (thus introducing systematic uncertainties on the measurement). In addition, we have only very limited possibilities to reject muons coming from semimuonic decays of pions and kaons, representing an important background both in single muon and dimuons analyses, in particular at low masses and low $\pt$. Finally, the degradation of the kinematics, imposed by the presence of the hadron absorber, plays a crucial role in determining the mass resolution for the resonances, especially at low masses.

~\\
To overcome these limitations, better exploiting the unique kinematic range accessible by the ALICE Muon Arm, the Muon Forward Tracker (MFT) was proposed in the context of the ALICE upgrade plans, to take place in the years 2017/2018 during the LHC shutdown. The MFT is a silicon pixel detector added in the Muon Spectrometer acceptance ($2.5 < \eta < 4$) upstream of the hadron absorber. The basic idea, motivating the integration of the MFT in the ALICE setup, is the possibility to match the extrapolated muon tracks, coming from the tracking chambers \emph{after} the absorber, with the clusters measured in the MFT planes \emph{before} the absorber; the match between the muon tracks and the MFT clusters being correct, muon tracks should gain enough pointing accuracy to permit a reliable measurement of their offset with respect to the primary vertex of the interaction. The measurement of the muons' offset should then allow one to: (i) disentangle prompt (quarkonia, thermal photons) from displaced (open HF) dimuons; (ii) distinguish open charm and open beauty dimuons on the basis of the analysis of the pairs' offset;
(iii) study HF via single muons down to $\pt~\approx~1~$GeV/$c$ with limited model dependence; (iv) study beauty production down to zero~$\pt$ via $J/\psi$ from $b$, a unique feature at the LHC in A-A collisions. In addition, applying quality cuts on the matching between the extrapolated tracks and the MFT clusters, it should be possible to reject a large fraction of background coming from semimuonic decays of primary and secondary pions and kaons, as well as punch-through hadrons arriving at the tracking chambers without being stopped in the absorber.

\section{MFT Design and Preliminary Performance Studies}

The MFT setup, as described in the simulation studies considered in the present report, is composed of five tracking planes placed at $z=50$, 58, 66, 74, 82~cm from the IP, before the hadron absorber and the scintillator interaction counter (VZERO) placed in front of it. Each plane is composed of a $0.2\,\%~x/X_0$ disk-shaped support element, and a 50~$\mu$m-thick assembly of silicon sensors and readout elements arranged in the front and back part of the support. Each plane thus contribute with $0.3\,\%~x/X_0$, leading to a total of $1.5\,\%~x/X_0$ for the whole MFT. The material budget traversed by the muons before arriving the MFT depends on the geometry of the beryllium beam pipe. In the simulations presented here we considered a cylindrical beam pipe having an internal radius of 2~cm and a thickness of 500~$\mu$m, a realistic scenario for the 2017 upgrade. It should be noted here that a cylindrical beam pipe induces at high rapidity multiple scattering effects not at all negligible; for this reason, present studies are on-going on possible setups with a modified conical beam-pipe. Conclusions on such investigations will be available in the next future, and within this report no comparison will be established between concurrent geometry setups of the beam pipe.

~\\
For the active elements covering the MFT planes we assumed a $20 \times 20~\mu$m$^2$ pixel segmentation, already available for a CMOS technology. Possibly occurring charge-dispersion effects may cause the activation of side pixels, in addition to the one actually traversed by the tracked particle: this is taken into account in the clusterization algorithm, which defines the center of the cluster averaging over the centers of the pixels composing it. In this way it is possible to recover a spatial resolution of the order of $20~\mu$m/$\sqrt{12}$. 


~\\
The tracking strategy starts from the muon tracks reconstructed after the hadron absorber. These are extrapolated back to the vertex region, taking into account both the energy loss and the multiple scattering induced by the hadron absorber. Each extrapolated track is then evaluated at the last plane of the MFT (the one closest to the absorber) and, for each cluster in this plane, its compatibility with the extrapolated track is evaluated in terms of the quantity:
\begin{equation}
 \chi_\mathrm{clust}^2 = \frac{ \Delta x^2 \sigma_y^2 +  \Delta y^2 \sigma_x^2 - 2 \cdot \Delta x \Delta y \, \mathrm{cov}(x,y) }
{ \sigma_x^2 \sigma_y^2 -  \mathrm{cov}^2(x,y) }~,
\end{equation}
where $\Delta x$ and $\Delta y$ represent the distance between the track position and the cluster along $x$ and $y$, while $\sigma_x^2$, $\sigma_y^2$ and $\mathrm{cov}(x,y)$ are the covariance matrix elements accounting for the combined uncertainty on the cluster and the track position. For each compatible cluster a new candidate track is created, whose parameters and their uncertainties are updated with the information given by the added cluster by means of a Kalman filter algorithm. Each candidate track is then extrapolated back to the next MFT plane, where a search for compatible clusters is performed in the same way as before. As the extrapolation proceeds towards the vertex region, the uncertainties on the parameters of the extrapolated tracks become smaller, the number of compatible clusters decreases and the number of candidate tracks converges. If more than one final candidate is found at the last extrapolation, the best one is chosen according to the global fit quality.

\begin{figure}[htbp]
  \vspace{-0.5cm}
  \includegraphics[width=.45\textwidth]{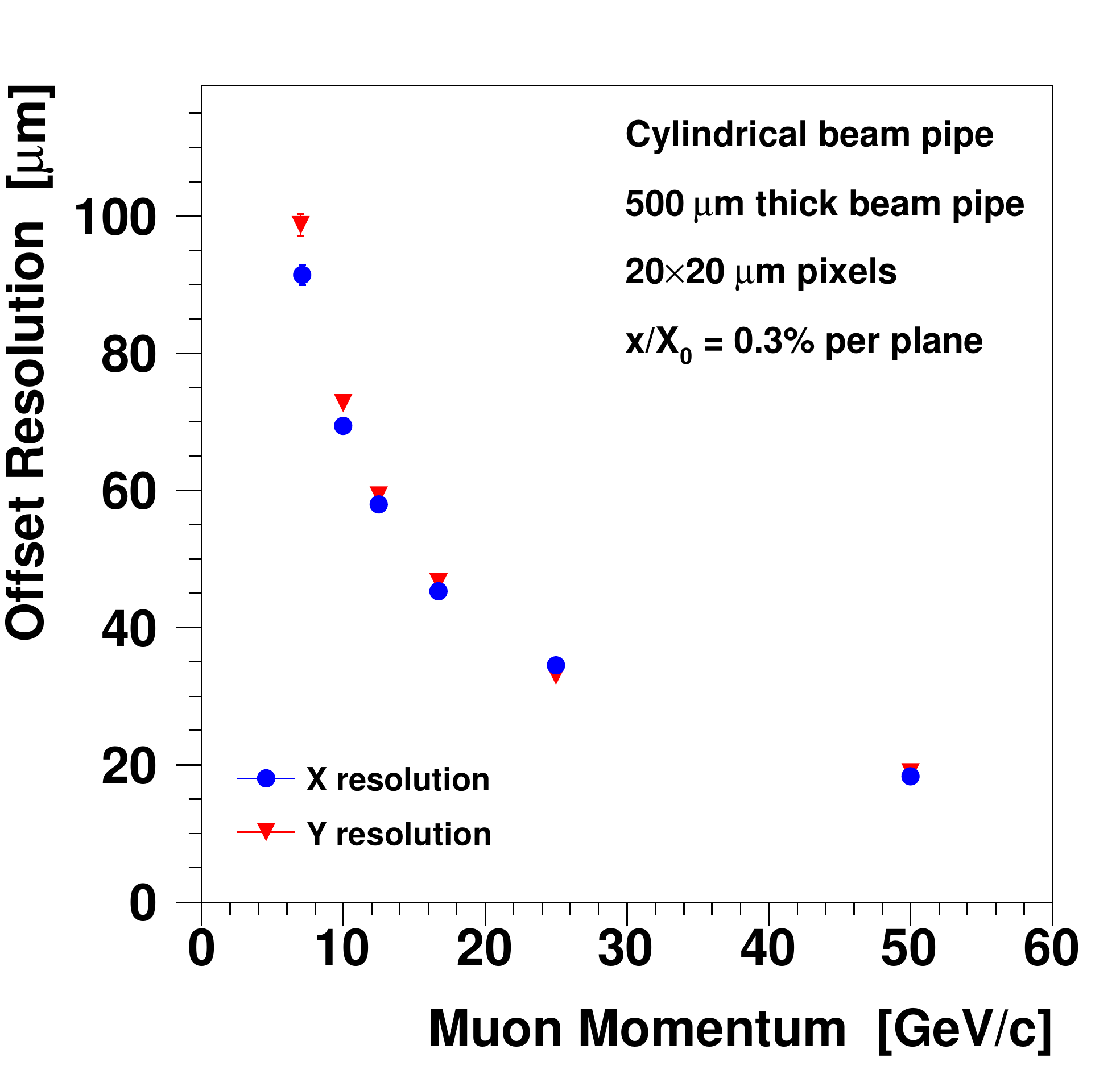} \hspace{.05\textwidth}
  \begin{minipage}[b]{.45\textwidth}
  \caption[\textwidth]{\label{fig:offsetResolution} Offset resolution for single muons as a function of the momentum, along the $x$ and $y$ directions. Error bars reflect the data sample available for each point. \vspace{0.05cm}}
  \end{minipage}
\vspace{-0.2cm}
\end{figure}

~\\
With a spatial resolution of the order of $\sim 20~\mu$m/$\sqrt{12}$ in the MFT planes, we have a reliable measurement of the muons' offset, i.e.~the transverse distance between the primary vertex and the muon track. To investigate the MFT performances in terms of the offset resolution for the single muons, single muons were generated at a fixed position in (0,~0,~0). The offset of the reconstructed muon tracks was then evaluated assuming the production vertex to be measured by the ALICE Internal Tracking System (ITS) with a precision of 100~$\mu$m along the $z$ direction. The observed offset resolution has been studied as a function of the muon momentum, see \figurename~\ref{fig:offsetResolution}. The resolution is as good as $\sim100~\mu$m even for muon momenta down to $\sim 7$~GeV/$c$. Slight differences between $x$ and $y$ directions probably reflect a remnant of the $x-y$ asymmetry introduced by the dipole magnetic field after the absorber. Studies are ongoing to characterize the offset resolution as a function of the muon rapidity, too, from which significant differences between a cylindrical and a conical beam pipe geometry could emerge.

\begin{figure}[htbp] 
\vspace{-0,4cm}
   \begin{center}
    \includegraphics[width=0.45\textwidth]{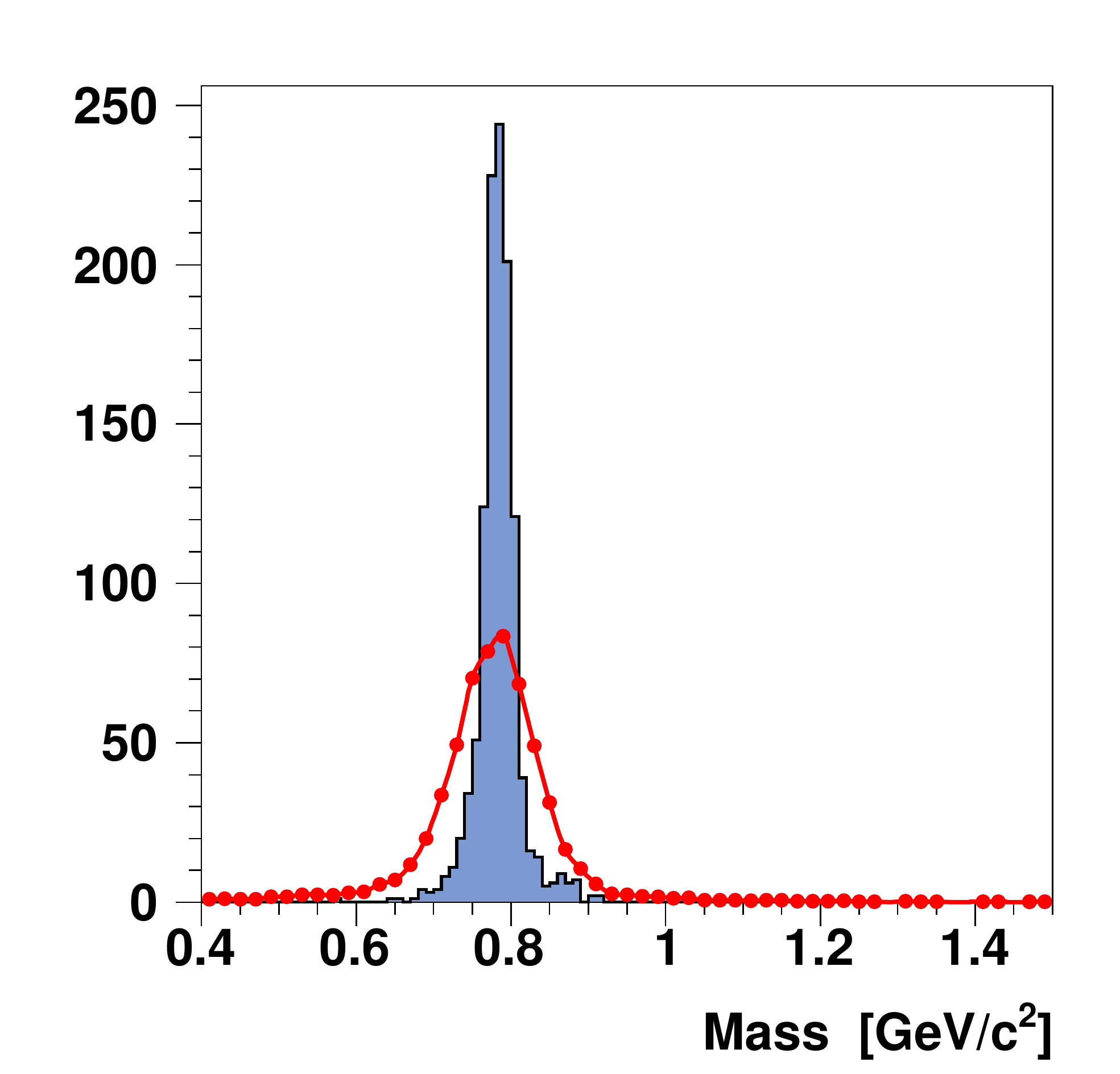} \hspace{0.05\textwidth}
    \includegraphics[width=0.45\textwidth]{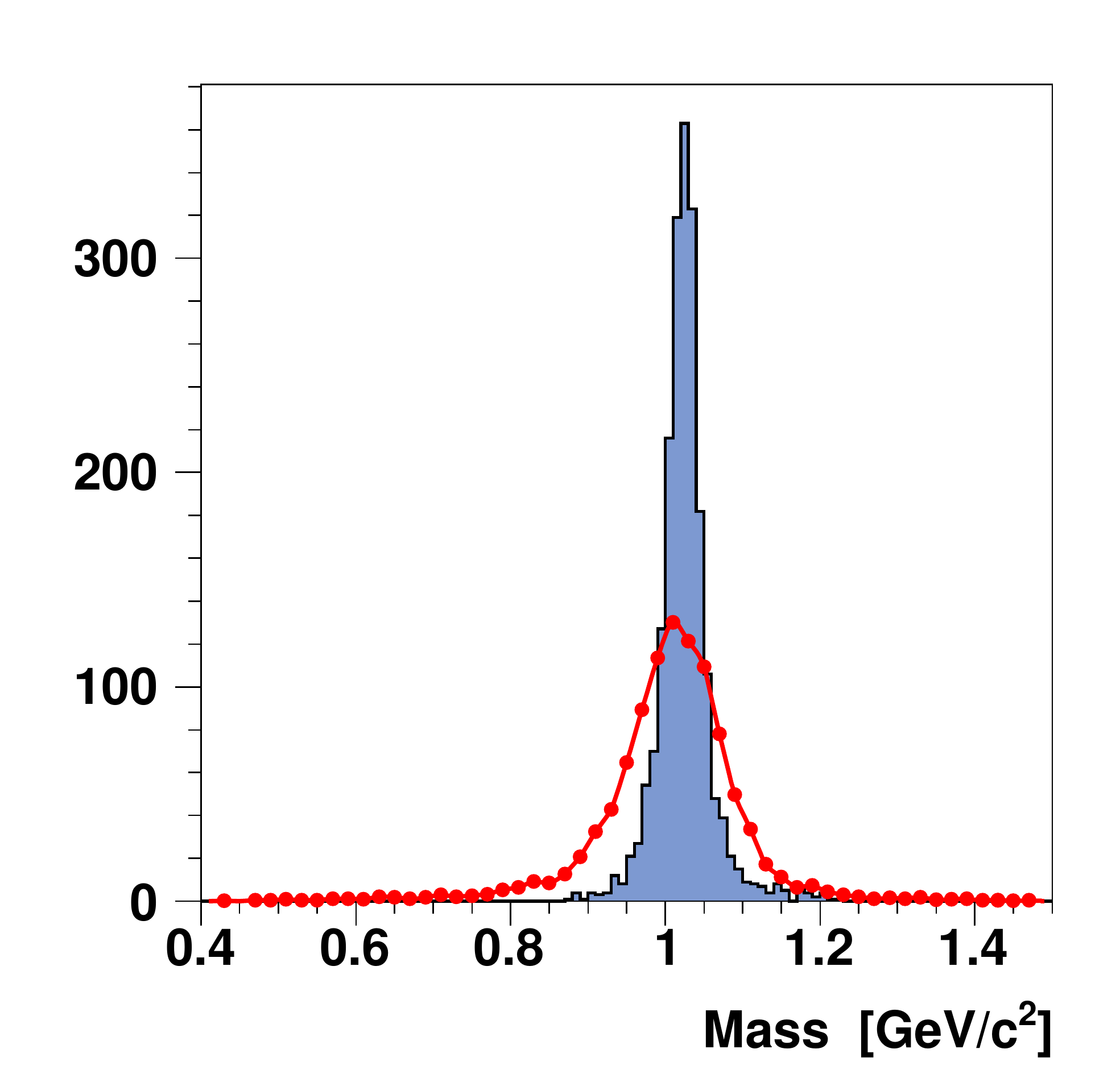}
    \end{center} 
\vspace{-0.2cm}
\caption[\textwidth]{Comparison between the mass resolution available with the current Muon Arm setup (red points and line) and the one achievable by means of the MFT (blue profile) for the $\omega$ (left panel) and $\phi$ (right panel) resonances.}
\label{fig:massResolution}
\vspace{0.1cm}
\end{figure}

\noindent The improved pointing accuracy gained by means of the matching between the extrapolated muon tracks and the MFT clusters, allows one to have a better evaluation of the opening angle for prompt muon pairs. This results in better mass resolutions for all the resonance measured in the dimuon channel, in particular for the lightest ones ($\eta$, $\omega$ and $\phi$) whose soft muons suffer more from the degradation of the kinematics induced by the hadron absorber. A comparison between the currently available resolution and the one resulting from the preliminary MFT simulations is shown in \figurename~\ref{fig:massResolution}, for the $\omega$ and $\phi$ mesons. With a simple Gaussian fit on the peaks obtained with the MFT simulations, one finds $\sigma_\omega \approx 16$~MeV/$c^2$ and $\sigma_\phi \approx 20$~MeV/$c^2$, an improvement up to a factor $\sim 3$ with respect to the resolutions available with the current Muon Arm setup.

\begin{figure}[htbp]
  \vspace{-0.2cm}
  \includegraphics[width=.60\textwidth]{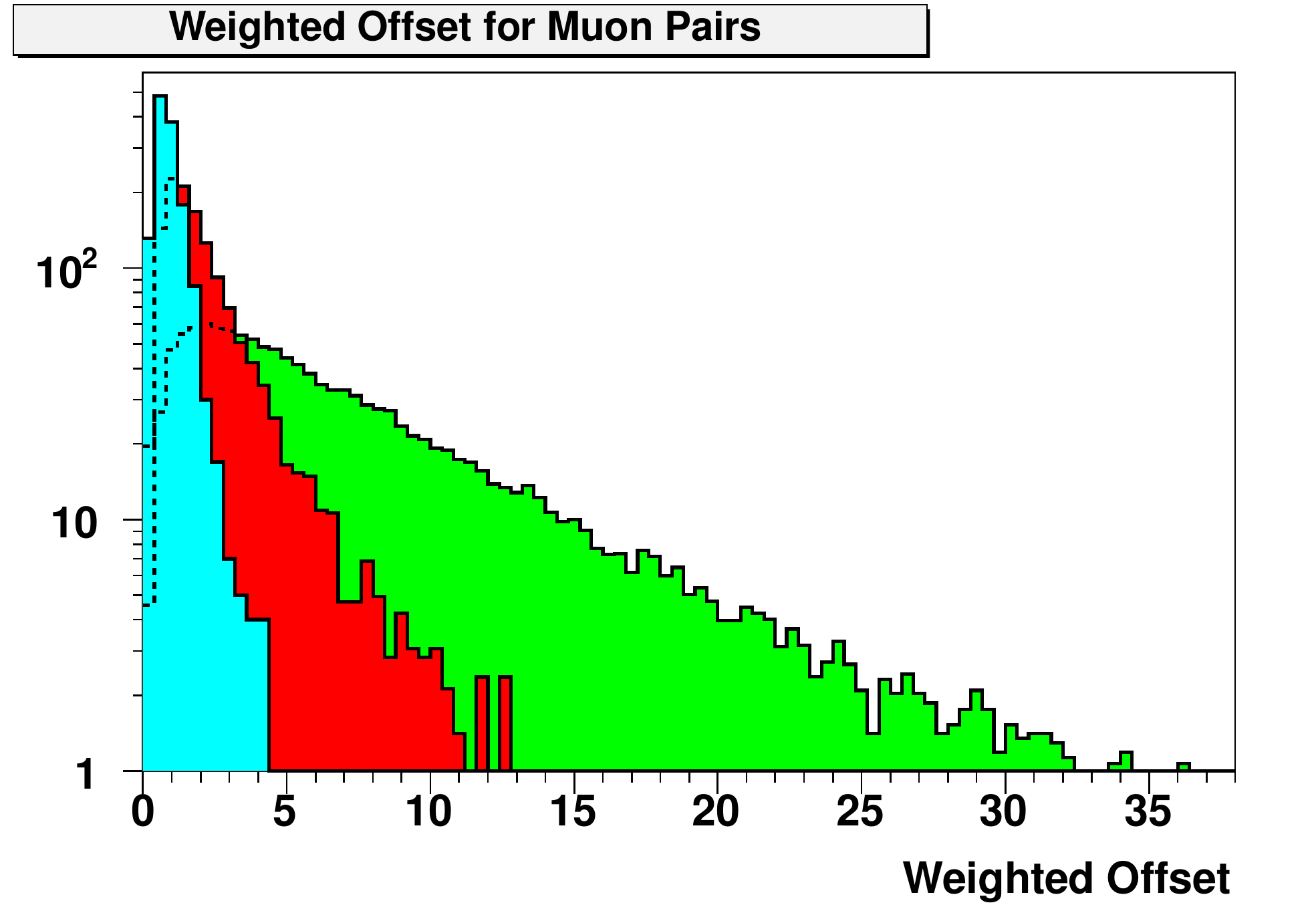} \hspace{.0\textwidth}
  \begin{minipage}[b]{.35\textwidth}
  \caption[\textwidth]{\label{fig:dimuOffset} Comparison between the weighted dimuon distributions for prompt (cyan profile), open charm (red profile) and open beauty (green profile) dimuons. \vspace{0.1cm}}
  \end{minipage}
\vspace{-1.cm}
\end{figure}

\noindent The last point addressed in these preliminary investigations on the MFT performances, concerns the possibility to distinguish prompt muon pairs from open charm and open beauty contributions on the basis of the analysis of the weighted offset of the dimuons, defined as $\Delta_{\mu\mu} = \big[0.5 \cdot (\Delta_{\mu1}^2 + \Delta_{\mu2}^2)\big]^{0.5}$, with $\Delta_{\mu} = \big[0.5 \cdot (\delta x^2 V_{xx}^{-1} + \delta y^2 V_{yy}^{-1} + 2\delta x \delta y V_{xy}^{-1})\big]^{0.5}$, where $\delta x$ and $\delta y$ are the $x$ and $y$ offset of the muon track, and $V^{-1}$ is the inverse of the covariance matrix accounting for the combined uncertainty on the track and the vertex position. From the definition above, we expect to find wider $\Delta_{\mu\mu}$ distributions as we go from prompt to displaced dimuons: this is exactly what we observe in \figurename~\ref{fig:dimuOffset}, where the prompt dimuons (in this case the $\phi$ meson) have a narrower distribution than the open charm dimuons ($c\tau \sim 150~\mu$m), which in turn have a narrower distribution than the open beauty dimuons ($c\tau \sim 500~\mu$m). This should permit to establish a reliable separation between the signal components, on a statistical basis, without any model dependence.

\section{Conclusions}

The muon physics program of the ALICE experiment is conditioned by the intrinsic limitations of the current setup of the Muon Arm. The addition of a silicon Muon Forward Tracker in the acceptance of the Muon Spectrometer should overcome these intrinsic limitations increasing the physics potential of the muon spectrometer, giving at the same time the possibility to perform new measurements of general interest for the whole ALICE physics. In this report we discussed the preliminary results of the simulations including the MFT, obtained in a low-multiplicity environment (matching efficiency $\sim 100\,\%$) with a focus on the offset resolution for single muons, the improvement of the mass resolution for the light mesons $\omega$ and $\phi$ and the weighted offset distributions for prompt and displaced muon pairs. These preliminary results are encouraging, and motivate an additional effort aiming to investigate the matching performances in high multiplicity, both as a function of the muons' kinematics and the MFT setup.

%
%
%
%

\bibliographystyle{ws-procs9x6}
\bibliography{biblio_URAS}

\end{document}